# Photophysics of blue quantum emitters in hexagonal Boron Nitride

Ivan Zhigulin[1,*], Karin Yamamura[1,*], Viktor Ivády[2,3,4], Angus Gale[1], Jake Horder[1], Charlene J. Lobo[1], Mehran Kianinia[1], Milos Toth[1,5], and Igor Aharonovich[1,5]

[1]School of Mathematical and Physical Sciences, University of Technology Sydney, Ultimo, New South Wales 2007, Australia
[2]Department of Physics of Complex Systems, ELTE Eötvös Loránd University, Egyetem tér 1-3, 1053 Budapest, Hungary
[3]MTA–ELTE Lendület "Momentum" NewQubit Research Group, Pázmány Péter sétány 1/A, 1117 Budapest, Hungary
[4]Department of Physics, Chemistry and Biology, Linköping University, Linköping, Sweden
[5]ARC Centre of Excellence for Transformative Meta-Optical Systems, University of Technology Sydney, Ultimo, New South Wales 2007, Australia
* These authors contributed equally.
Corresponding author igor.aharonovich@uts.edu.au

**Abstract**

Colour centres in hexagonal boron nitride (hBN) have emerged as intriguing contenders for integrated quantum photonics. In this work, we present detailed photophysical analysis of hBN single emitters emitting at the blue spectral range. The emitters are fabricated by different electron beam irradiation and annealing conditions and exhibit narrow-band luminescence centred at 436 nm. Photon statistics as well as rigorous photodynamics analysis unveils potential level structure of the emitters, which suggests lack of a metastable state, supported by a theoretical analysis. The potential defect can have an electronic structure with fully occupied defect state in the lower half of the hBN band gap and empty defect state in the upper half of the band gap. Overall, our results are important to understand the photophysical properties of the emerging family of blue quantum emitters in hBN as potential sources for scalable quantum photonic applications.

**Introduction**

Single photon emitters (SPEs) are widely acknowledged as key enablers to establish and deploy quantum communication and computing, which involves on-demand generation of high purity single photon emission[1-3]. Hexagonal boron nitride (hBN) has gained attention due to its unique properties of wide layer-dependent bandgap centred around 6 eV, high exciton binding energies, presence of optically active spin-defects and capability to host room-temperature (RT) bright SPEs[4-11]. hBN is also attracting attention for its use as an emerging optoelectronic material for the deep ultraviolet range[12].

Recently, color centres in hBN emitting at the blue spectral range, termed 'blue emitters', were discovered by cathodoluminescence (CL) measurements [13]. This group of emitters generally displays ultra-bright, spectrally stable and narrowband emission with a zero phonon line (ZPL) consistently centred around 436 nm [13, 14]. It was shown that these defects are closely related to the presence of a signature UV emission at 4.1 eV [9, 14-16]. Pre-irradiation of hBN, such as high temperature annealing in a nitrogen atmosphere, results in higher yield of the signature UV emission and thus higher numbers of blue colour centres [15]. Additionally, at cryogenic temperatures, these defects have stable emission with sub-GHz linewidths and minimal spectral diffusion compared to other quantum emitters in hBN [15]. Very recently, two-photon interference was demonstrated, opening new opportunities for 2D materials in optical quantum information [17].

To date, the crystallographic origin of the blue emitters is still under debate, and likewise, little is known about the photophysical nature of the defects. It is thus important to investigate the properties and characteristics of this group of defects to gain insight into their formation mechanisms as well as into their optical properties before they can be fully utilised for practical applications.

Here, we present detailed photophysical analysis of the blue SPEs in hBN created using focused electron beam irradiations with varying dosages. We investigate photostability, saturation behaviour, lifetime and transition rates employing time resolved excitations and time correlation measurements. We use these results to propose the electronic structure of blue SPEs and the position of their energy levels relative to the hBN valence and conduction bands.

We examined in detail over ten blue SPEs found in three different hBN flakes. Although electron beam irradiation conditions for each flake varied, the obtained results (in terms of the ZPL emission wavelength) remained consistent, supporting the hypothesis that the blue quantum emitters have the same defect origin. For clarity, we therefore introduce an emitter labelling scheme (eg. flake 1, emitter 1 = F1E1), which will be used throughout this manuscript.

We start with analysing the optical properties of emitters, formed by electron irradiation (schematically shown in figure 1a). First, a spectrum of each emitter was acquired and the key parameters - namely emission at saturation, excited state fluorescence lifetime, $\tau_1$, and autocorrelation, $g^{(2)}(\tau)$, were measured, as shown in Figure 1 for F3E1. Spectral characteristics of emitters were measured using a continuous-wave (CW) 405 nm laser at powers ranging from 0.66 mW to 5.00 mW. The photoluminescence (PL) emission of F3E1 at 0.66 mW and 5.00 mW centred at 436 nm is shown in Figure 1b.

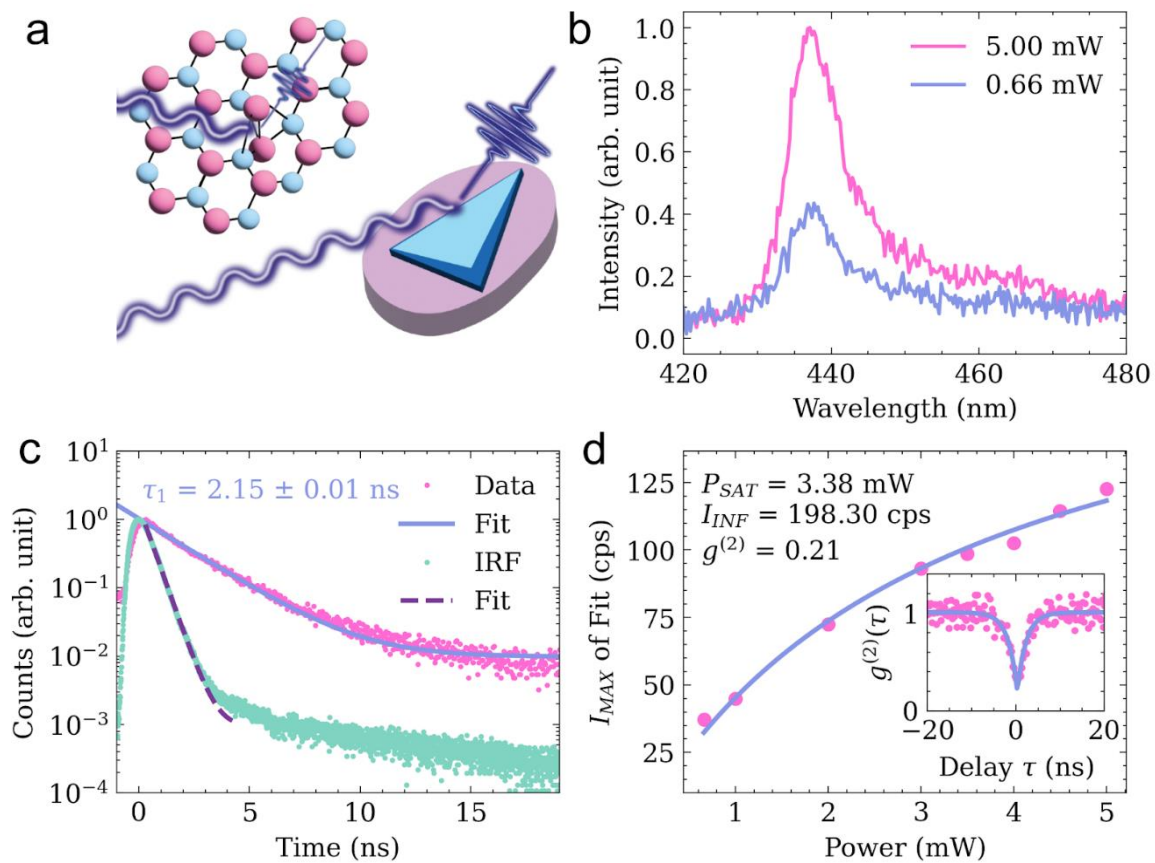

***Figure 1****. Spectral characteristics of emitter 1 in flake 3 (F3E1). (a) Diagram of hBN lattice and flake. (b) Photoluminescence spectra of the emitter excited with a 405 nm CW laser at powers of 0.66 mW (shown in purple) and 5.00 mW (shown in pink) at room temperature. (c) Emission lifetime measurement of the blue emitter excited with a 402 nm pulsed laser at 0.66 mW and frequency of 30 MHz. A lifetime $\tau_1 = 2.15 \pm 0.01$ ns was deduced from an exponential fit (blue line). Green dots show the instrument response function (IRF) with a corresponding fit that extracts a value of 0.475 ± 0.005 ns. (d) Power saturation measurement of the same emitter. Blue line shows the fit of the experimental data (pink dots) using the power dependence model (shown in Equation (1)). Inset in (d) shows autocorrelation function $g^{(2)}(0)$ of 0.21 proving the single photon nature of this emitter.*

To study the emission lifetimes, a 402 nm pulsed laser at 0.66 mW and frequency of 30 MHz was employed. All lifetime measurements were conducted at room temperature (RT) under ambient conditions. Figure 1c shows the emission lifetime $\tau_1 = 2.151 \pm 0.009$ ns of F3E1, obtained by fitting the data with a single exponential decay. The value is consistent with lifetimes measured for other blue emitters found in another flake, 2.019 ± 0.003 ns for F2E1, 1.931 ± 0.003 ns for F2E3, and 2.160 ± 0.017 ns for F2E8. Thus, the lifetime of ~ 2 ns is consistent with other studies of these emitters [13-15, 18].

To gain the saturation information, PL spectra were acquired at each power for 3 seconds. The maximum value of each spectrum was used as a point in the saturation plot, and the saturation power ($P_{SAT}$) and the maximum obtainable PL intensity at saturation ($I_{INF}$) were then extracted from a fit to the data using Equation (1):

$$I(P) = \frac{I_{INF} \cdot P}{P_{SAT} + P} \qquad (1)$$

Figure 1d shows the fluorescence saturation curve of F3E1 with a calculated value of 3.38 mW for $P_{SAT}$. Measured $P_{SAT}$ values of other blue emitters are 1.50 mW for F2E1, 2.10 mW for F2E4, 2.40 mW for F2E5, 6.50 mW for F2E8, and 8.90 mW for F2E9. Thus, examined emitters have $P_{SAT}$ ranging from 1.50 mW to 8.90 mW, similar to previously reported for blue emitters in hBN at room temperature, which are slightly above typical saturation values for the visible (~ 2 eV) emitters in hBN [14, 15]. Such relatively high $P_{SAT}$ could be attributed to absorption cross-section of blue emitters from 405 nm excitation. The inset in Figure 1d shows the autocorrelation function at zero delay time, $g^{(2)}(0)$ of 0.21 verifying the single photon nature of this emitter.

Next, we focus on the photostability and photon statistics of the blue emitters, to understand their energy levels. Saturation power plays a key role in these measurements, because it is necessary to saturate a system in order to observe an increasing bunching effect in $g^{(2)}(\tau)$ measurements. The relatively high saturation power of blue SPEs makes this measurement difficult, as it requires prolonged excitation under high power, which causes bleaching in some of the emitters. We observed repeatedly that exciting a single blue emitter for > 4 hours at powers greater than 1.00 mW causes the system to go irreversibly into its dark state.

Figure 2a shows $g^{(2)}(\tau)$ measurements of four blue emitters found in Flake 1 acquired between 0.69 - 2.00 mW for over 4 hours. All emitters have a $g^{(2)}(0)$ value below 0.5 indicating that they are SPEs. Only slight bunching was observed from the emitters above 1.00 mW, which becomes marginally more noticeable around 2.00 mW. Note that at longer delay times the line is flat, indicating that no multiple shelving states are present [19-23]. This photophysical behaviour is substantially different from the majority of defect based SPEs in diamond, silicon carbide, or even hBN that often exhibit strong bunching with multiple obvious decay pathways [20, 22, 24]. These various decay channels (evident by changes in the slope of the $g^{(2)}(\tau)$ at longer delay times) are often a signature of a metastable state or recurring blinking behaviour[25]. These effects are nearly negligible for the blue emitters that behave as a nearly ideal two-level system, as will also be discussed below. Indeed, $g^{(2)}(\tau)$ curves with high $P_{SAT}$ above 5.00 mW (approximately ~ 25% of emitters) showed no bunching at the measured powers. Note, that many of the emitters were eventually bleached during the $g^{(2)}(\tau)$ acquisition at excitation powers higher than 2.00 mW after approximately one hour.

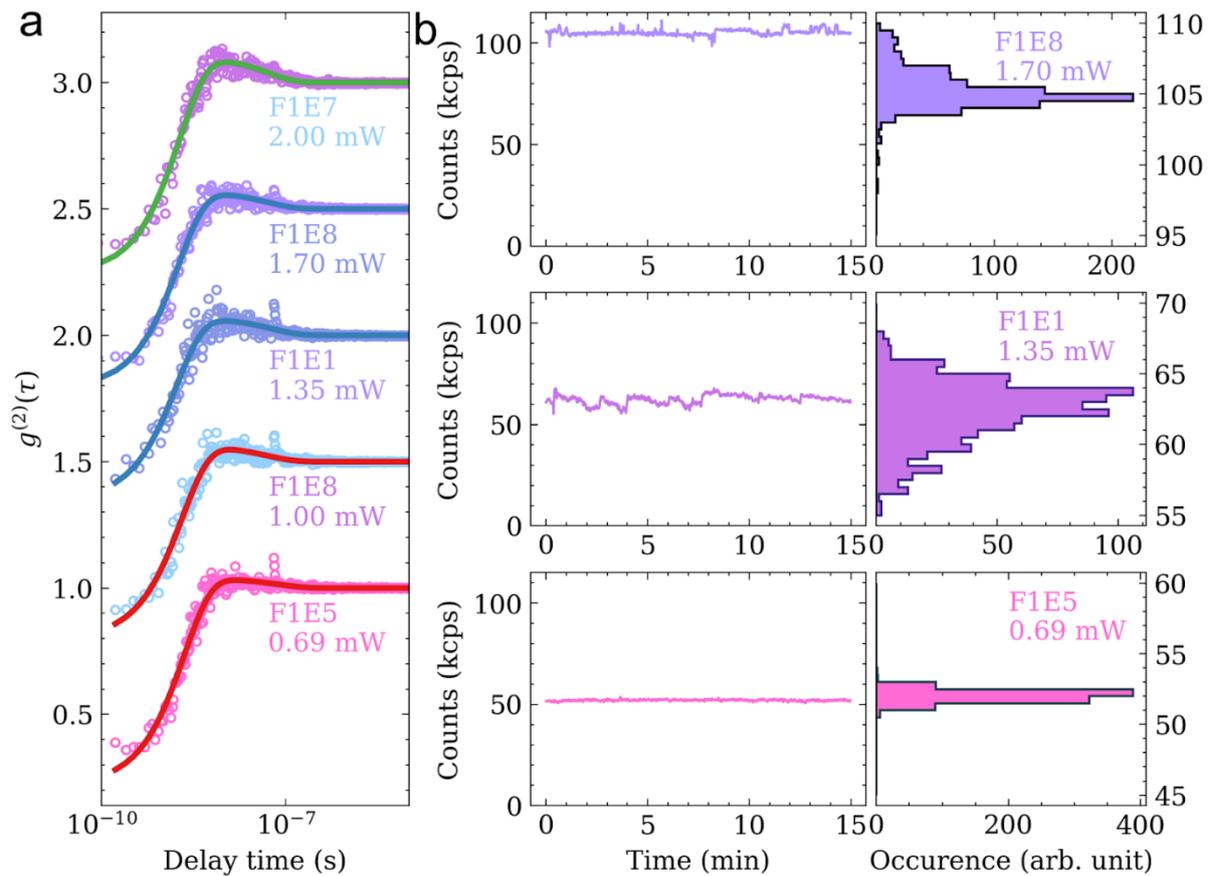

*Figure 2*. Photostability and photodynamics analysis of blue SPEs. (a) Power-dependent l $g^{(2)}(\tau)$ measurements at 0.69 mW, 1.00 mW, 1.35 mW, 1.70 mW and 2.00 mW on four different emitters on Flake 1. (b) Timetraces with 1 second resolution and corresponding histograms of fluorescence intensity for three emitters. Range of histograms is kept constant at 15 kilo counts/second.

The luminescence timetraces (kilo counts/sec) of emitters F1E5, F1E1 and F1E8 at excitation powers of 0.69 mW, 1.35 mW and 1.70 mW are shown in Figure 2b. Timetraces of each of these emitters are stable over time, without obvious blinking or fluorescence intermittency. The ranges of distribution of the photon counts are within 10% of the mean, for each emitter. This is important, as it suggests that the emitters that remain optically active, exhibit excellent photostability over long excitation periods [19, 22, 26, 27].

To investigate the rates quantitatively, we recorded the autocorrelation measurement as a function of excitation power [19, 20, 22]. The $g^{(2)}(\tau)$ function taken at various excitation powers can be fit with a standard bi exponential function using Equation (2):

$$g^2(\tau) = 1 - (1+a)e^{\frac{-|\tau|}{\tau_1}} + ae^{\frac{-|\tau|}{\tau_2}}$$

(2)

Where $\tau_1$, $\tau_2$ correspond to lifetimes of antibunching and bunching respectively, while *a* is a parameter that describes bunching amplitude.

Figure 3a shows power-dependent plots of $g^{(2)}(\tau)$ measurements of F3E1, with the value of $g^{(2)}(0)$ remaining below 0.5 at each excitation. Figure 3b,c show the power-dependent parameters of $\tau_1$, $\tau_2$ and the bunching factor *a* obtained from fitting Equation (2) to the $g^{(2)}(\tau)$ measurements.

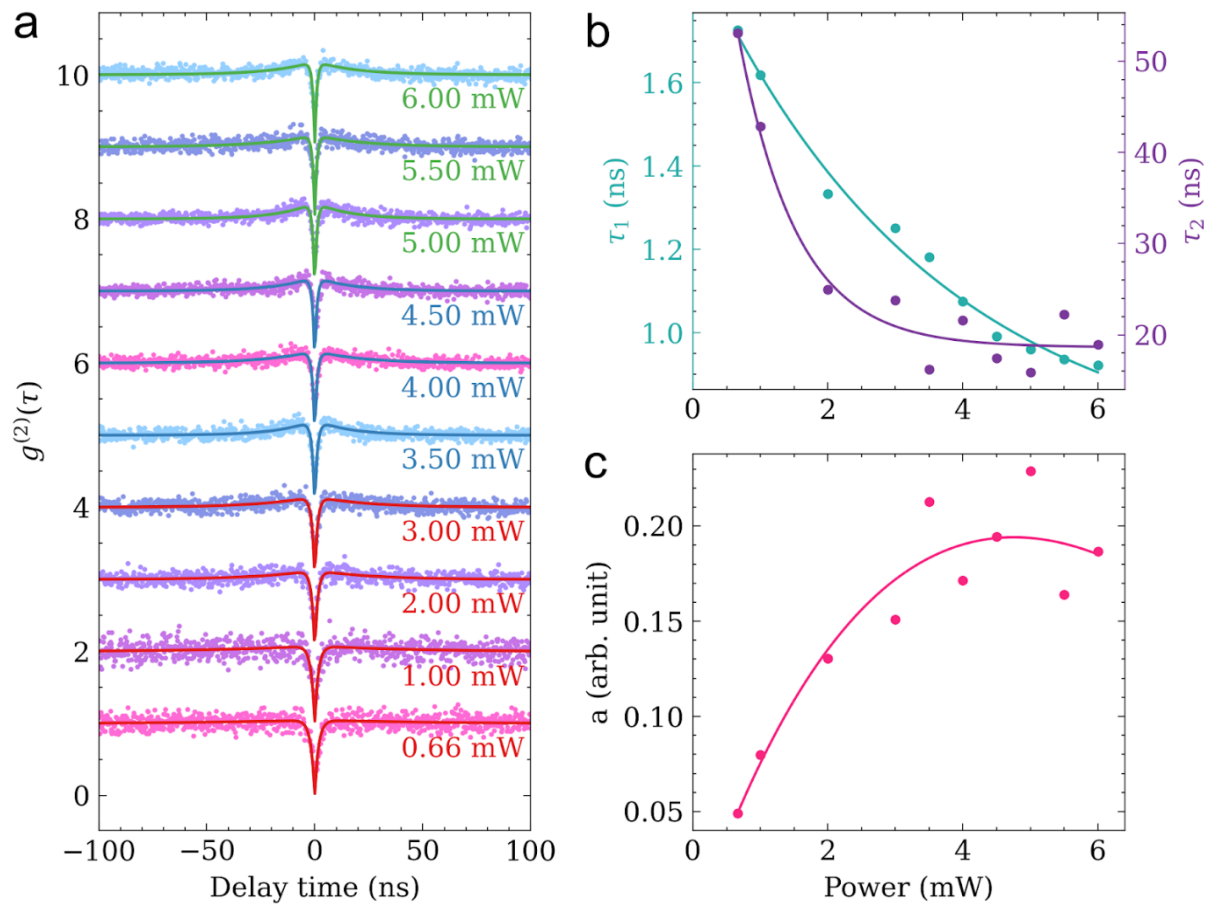

*Figure 3. Power dependent analysis of energy levels of F3E1. (a) Second-order autocorrelation measurements taken at various excitation powers. $g^{(2)}(\tau)$ traces are normalised and stacked vertically. (b) Lifetime $\tau_1$ of transitions from the excited state to the ground state (green dots) and metastable state lifetime $\tau_2$ (purple dots) as a function of laser power. (c) Pink dots show the bunching factor, a, vs excitation power.*

From the fitting of $\tau_1$, $\tau_2$ and a, we calculated the transition rates according to the three-level rate equation [22, 23]. The transition rates between the three states were calculated by measuring the power-dependent $g^2(\tau)$. The excitation rate from the ground state to the excited state, radiative decay rate from the excited state to the ground state, decay rate from the excited state to the metastable state, and decay rate from the metastable state to the ground state are expressed as $k_{12}$, $k_{21}$, $k_{23}$, and $k_{31}$, respectively. The $\tau_1$, $\tau_2$ and a can be derived using Equations (3) and (4). The value of $k_{12}$ depends on the power, and $k_{21}$ and $k_{31}$ are assumed to be power-dependent states. The transition rates were obtained following a combination of Equations (3)-(9).

$$a = \frac{1 - \tau_2 k_{31}}{k_{31}(\tau_2 - \tau_1)} \tag{3}$$

$$\tau_{1,2} = \frac{2}{A \pm \sqrt{A^2 - 4B}} \tag{4}$$

$$k_{31} = \frac{d \cdot P}{P + C} + k_{31}^0 \tag{5}$$

$$k_{31}^0 = \frac{1}{\tau_2^0} \tag{6}$$

$$d = \frac{\frac{1}{\tau_2^\infty} - (1 + a^\infty)\frac{1}{\tau_2^0}}{a^\infty + 1} \tag{7}$$

$$k_{23} = \frac{1}{\tau_2^\infty} - k_{31}^0 - d \tag{8}$$

$$k_{21} = \frac{1}{\tau_1^0} - k_{23} \tag{9}$$

Where, $A = k_{12} + k_{21} + k_{23} + k_{31}$ and $B = k_{12}(k_{23}+k_{31}) + k_{31}(k_{21}+k_{23})$, P is the excitation power, d and C are the coefficients related to the saturation behaviour. The rate per power coefficient, σ, is determined from fitting $\tau_1$ [22]. The same procedure was repeated for various emitters, as shown in table 1 below where the calculated transition rates are listed.

| Emitter | $k_{21}$ (MHz) | $k_{23}$ (MHz) | $k_{31}^0$ (MHz) | σ (MHz/mW) |
|---|---|---|---|---|
| F3E1 | 337.6 | 3.5 | 10.7 | 121.1 |
| F1E12 | 323.9 | 41.3 | 24.6 | 87.4 |
| F2E1 | 344.4 | 20.7 | 31.7 | 137.0 |
| F2E4 | 365.5 | 17.4 | 22.0 | 145.5 |
| F2E5 | 312.0 | 23.5 | 15.7 | 183.0 |

*Table 1. List of the calculated transition rates of $k_{21}$, $k_{23}$, $k_{31}^0$ and σ of F3E1, F1E12, F2E1, F2E4, and F2E5.*

The calculated transition rates are $k_{21}$ = 337.6 MHz, $k_{23}$ = 3.5 MHz, $k_{31}^0$ = 10.7 for the specific emitter F3E1. Similar order of magnitude results were also obtained for the other emitters. The excited state transition rates are an order of magnitude faster than the transitions to/from the metastable state, $k_{23}/k_{31}^0$. These results are consistent with the fact that the blue emitters behave as a nearly ideal two-level system.

We can also comment on the lower limit of the quantum efficiency of the blue emitters. Note that precise measurement of quantum efficiency is challenging as it requires modification of local dielectric environment[28, 29]. Most of the investigated emitters yielded a count rate of ~ 500 kHz, at saturation, with ~ 2.1 ns (~ 450 MHz) fluorescence lifetime. The setup at 440 nm is highly inefficient, mostly driven by the low internal quantum efficiencies of the avalanche photodetectors (~ 10%). Hence, assuming an upper collection efficiency of our confocal setup at 4% in the visible (~ 700 nm) range, the efficiency will drop to below 0.5% at the blue spectral range. This puts a lower bound of the quantum efficiency at ~ 20%, which makes it a bright solid state quantum system.

Finally, we discuss the potential level structure and excitation cycle for the blue SPEs. Short wavelength excitations, e.g. the 405 nm laser used in the experiments, may change the charge state of defects and turn them into a dark state. Such photoionization processes can induce either blinking or loss of the emitter. The reported robustness of the blue emission under continuous excitation at moderate powers suggests that the bright charge state of the underlying defect structure is stable and no photoionization processes taking place. Likewise, no further fluorescence lines at longer wavelengths were observed, indicating lack of photochromism [23, 30-32]. This observation can be used to narrow down the possible electronic structures giving rise to the blue emission.

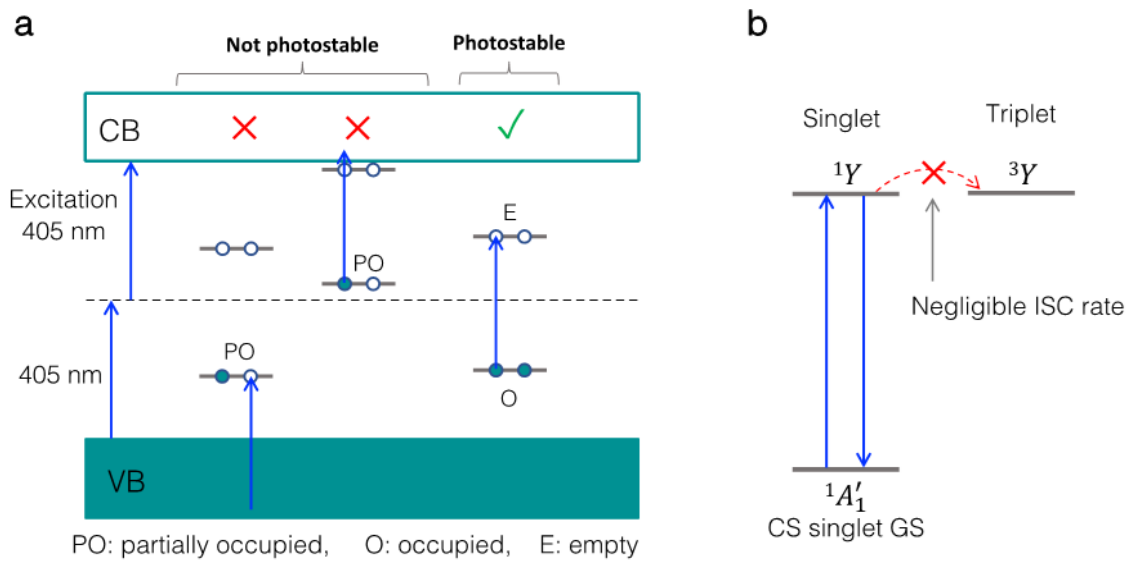

*Figure 4. Tentative electronic structure of the blue emitter. (a) Stable and unstable electronic configuration under 405 nm excitation. Defects with fully occupied defect states in the lower half of the band gap and empty defect states in the upper half of the band gap are photostable. (b) Many-body electronic structure of the photostable charge configuration with a closed shell (CS) singlet ground state (GS) of $^1A_1'$ symmetry. The triplet state decouples from the singlets and it does not interfere with the optical emission.*

To evaluate the photostability of defect electronic structures, we need to take valence band-to-defect and defect-to-conduction band transitions into consideration. As illustrated in Fig. 4a, the energy of the 405 nm photon is approximately half of the band gap of hBN. Therefore, partially occupied electronic states appearing in the lower half of the band gap are not photostable, an electron from the valence band can be excited to the defect state and change the charge state of the defect. Following the same argument, partially occupied states in the upper half of the band gap are neither photostable. The electron on the defect state can be excited to the conduction band and change the state again.

From the reported photostability of the blue emitter, we deduce that the most likely electronic structure of the responsible defect includes fully occupied defect state(s) in the lower half of the band gap and empty defect state(s) in the upper half of the band gap, see Fig. 4a. In this case the many-body ground state of the defect is a singlet that transforms according to the trivial representation of the corresponding point group. For $D_{3h}$ symmetry the ground state is $^1A_1'$. The optically excited state is also a singlet with an orbital symmetry "Y" that depends on the symmetry of the partially occupied single particle defect states in the excited states of the defect. Next to the singlet excited state there must be a triplet excited state, which can be obtained by flipping one of the electrons on the partially occupied defect states in the $^1Y$ excited state configuration, as illustrated in Fig. 4b. Since the orbital symmetry does not change, the triplet state can be labelled as $^3Y$. Transition from this state to the ground state is spin forbidden. Furthermore, since the orbital state of the two excited states is the same, spin-orbit coupling between the singlet and triplet states is also forbidden in first order. Therefore, the triplet excited state decouples from the optical processes and does not interfere with the fast radiative decay of the blue emitter, in accord with lack of strong bunching and slow rates to/from the metastable state. The electronic structure of split interstitial defects fulfils the criteria deduced from the photostability of the blue emitter [33]. However, to identify the microscopic structure of the blue emitter comprehensive first principles studies and comparison of theoretical and experimental results are needed.

In summary, we performed detailed photophysical analysis of the blue SPEs in hBN. Majority of the emitters exhibit excellent photostability and behave like an ideal two-level system without significant bunching in the autocorrelation function. The ZPLs of the emitters is consistently 436 nm and the lifetime of the blue SPEs is ~ 2 ns. Based on our excitation dynamics, the defects' ground and excited state are located at the middle of the hBN bandgap, which is advantageous for stability and inhibition of ionisation. Our results provide further insights into emerging class of hBN SPEs, that are promising for scalable quantum applications.

**Acknowledgments**

This work is supported by the Australian Research Council (CE200100010, FT220100053) and the Office of Naval Research Global (N62909-22-1-2028). The authors thank the ANFF node of UTS for access to facilities. V.I acknowledges support from the National Research, Development, and Innovation Office of Hungary (NKFIH) (Grant No. FK 145395), the Ministry of Culture and Innovation and the National Research, Development and Innovation Office within the Quantum Information National Laboratory of Hungary (Grant No. 2022-2.1.1-NL-2022-00004), and the Knut and Alice Wallenberg Foundation through WBSQD2 project (Grant No. 2018.0071).